\documentclass[12pt]{iopart}
\usepackage{graphicx}

\usepackage{color,ulem} 
\usepackage{amssymb}
\usepackage{iopams}  

\begin{document}



\newcommand{\FELIPE}[1]{\textcolor{blue}{\fbox{Felipe} {\sl#1}}}
\newcommand{\ch}[1]{\textcolor{green}{\fbox{PAMN} {\sl#1}}}
\newcommand{\Rfaqat}[1]{\textcolor{red}{\fbox{RA} {\sl#1}}}
\newcommand{\Rafael}[1]{\textcolor{black}{#1}}

\newcommand{\be}{\begin{equation}}
\newcommand{\ee}{\end{equation}}
\newcommand{\beq}{\begin{eqnarray}}
\newcommand{\eeq}{\end{eqnarray}}
\newcommand{\bea}{\begin{array}}
\newcommand{\eea}{\end{array}}

\title[]{Gain-assisted optical tweezing of plasmonic and large refractive index microspheres }

\author{R.  Ali$^{1,*}$, R. S. Dutra$^2$,  F. A. Pinheiro$^3$  \& P. A. Maia Neto$^3$}
\address{$^1$ Applied Physics Department, Gleb Wataghin Physics Institute, University of Campinas, Campinas 13083-859, SP,
Brazil}
\address{$^2$ {LISComp-IFRJ, Instituto Federal de Educa\c{c}\~ao, Ci\^encia e Tecnologia, Rua Sebasti\~ao de Lacerda, Paracambi, RJ, 26600-000, Brasil}l}
\address{$^3$ Instituto de F\'isica, Universidade Federal do Rio de Janeiro, Caixa Postal 68528, Rio de Janeiro, RJ, 21941-972, Brasil
}
\ead{$^*$rali.physicist@gmail.com}


\begin{abstract}
We have theoretically investigated optical tweezing of gain-functionalized microspheres using a highly focused single beam in the nonparaxial regime. 
 We employ the Mie-Debye theory of optical tweezers to calculate the optical force acting on homogeneous and core-shell Mie microspheres with gain. We demonstrate that the optical gain plays a crucial role in optical manipulation, especially to optimize the restoring force and thus allowing for trapping of large refractive index and  plasmonic particles.  Indeed we demonstrate that one can trap such particles, which is usually not possible in the case of passive media, by functionalizing them with an optical gain material.  We show that by varying the value of the gain, which can be realized by changing the pump power, one can not only achieve trapping but also manipulate
  the equilibrium position of the tweezed particle. Altogether our findings open new venues for gain-assisted optomechanics, where gain functionalized systems could facilitate optical trapping and manipulation of plasmonic nanoparticles in particular, with potential applications in self-assembling of nanoparticle suspensions and on a chip.  
\end{abstract}

\section{Introduction}

Since the invention of optical tweezers~\cite{ashkin1986,Ashkin1987}, optical manipulation have experienced notable progress, both scientifically and technologically, with many applications in several areas such as atomic physics, optics and biology~\cite{Ashkin2006,Polimeno2018}. Optical manipulation comprises not only trapping, but also cooling, binding, sorting and transporting~\cite{gao2017}. These tasks require controlling both linear and angular momenta of light to achieve tractor beams~\cite{sukhov2011}, pulling forces~\cite{chen2011,dogariu2013,Ali2020b,Ali2021}, and optical torques at the micro and nanoscale~\cite{Grier2003,Grier2012,Diniz2019,Ali2020}, unveiling new aspects of light-matter interactions.

In typical experimental setups of optical tweezers, a single tightly-focused laser beam is used to control the position of a (typically spherical) particle. 
The exchange of linear momentum upon light reflection by the particle produces a pushing, radiation pressure force. On the other hand, the refraction of a strongly-focused light beam gives rise to a restoring force bringing the particle to the focal point. Stable trapping is achieved at the point at which the two force components equilibrate~\cite{Ashkin1992}.
The detrimental radiation pressure effect tends to be dominant in the case of large refractive index, absorbing, and resonant particles.  As a result trapping is typically difficult to implement in those cases. 
Plasmonic particles of sizes larger than the wavelength are important examples as in this case 
even large numerical aperture (NA) objectives cannot provide  restoring forces sufficiently large to compensate for the strong radiation pressure component~\cite{min2013}. 
Despite challenging, trapping of large metallic particles remains a sough-after effect for practical applications that include spectroscopy, catalysis, micro-fabrication, biotechnology, and medical science~\cite{zhang2018}.

In order to circumvent these limitations, several techniques have been proposed to enhance restoring forces. For instance, optical tweezing can be  facilitated and optimized by using anti-reflecting coatings~\cite{Craig2015,Wang2018}. Additional proposals include
 suppressing  backscattering using composite media to achieve the Kerker condition~\cite{Ali2018}, tailoring the optical properties of the surrounding host medium \cite{Marago2008,Ali2020c},  exciting surface plasmon polaritons in the particle~\cite{Volpe2006,Min2013,Brzobohaty2015}, and exploring
 the optical response of chiral particles illuminated by circularly-polarized beams~\cite{Ali2020josab}.  All of the aforementioned works  consider passive media only. 

However, recent progress in nanofabrication techniques has opened the way for the design of gain-functionalized plasmonic nanoparticles, where encapsulated dye molecules are doped in homogeneous and coated metallic spheres, leading not only to loss compensation but also to a strong resonant coupling between the gain medium and the metal~\cite{deluca2012}.    
Recent applications of optical manipulation assisted by gain media include tailored optical forces by dual-beams on gain enriched Rayleigh particles~\cite{Pezzi2019,Polimeno2020} and optical pulling forces \cite{Bian2017}.  Advances in the field of metamaterials also allow to design  microspheres with optical gain \cite{Boardman2011,Campione2011}, with several applications \cite{Shen2017,Trigo2020,Mizrahi2010}. 

In this paper, we investigate optical tweezing of gain-funcionalized spherical particles, including core-shell plasmonic structures, using a single tightly focused laser beam.  
 We demonstrate that optical gain plays a crucial role to facilitate and optimize trapping {of plasmonic and large spheres with high refractive indexes, which
  cannot be trapped by a single focused beam without the assistance of gain}.  
  We systematically show that a moderate amount of  gain allows for stable trapping. We consider three 
  different realistic structural morphologies: homogeneous microspheres and core-shell particles  with either a dielectric or plasmonic core and a  dye-doped shell.
 For each configuration, we  identify the corresponding lowest and the highest value of gain that differentiate the regions where optical trapping and optical pulling {occur. We also demonstrate how the trapping position depends on the gain value, which can be adjusted dynamically by varying the pump power.}

{\section{Methodology }} 
 
In a  typical optical tweezers setup, a single incident laser beam is strongly focused by a high numerical aperture (NA) objective lens producing a non-paraxial beam  in the sample region, which contains an aqueous suspension (refractive index $n_w=\sqrt{\epsilon_w}$) of spherical particles. 
For simplicity, we consider an aplanatic (aberration-free) trapping beam.
 An extension to include optical aberrations can be implemented along the lines of Ref.~\cite{Dutra2014}.
The electric field representing the focused laser beam is written as a superposition of plane waves as follows~\cite{RichardsWolf}
\beq \label{RW}
\mathbf{E}_{\rm in}(\mathbf{r}) &&\hspace{-10pt}= {E}_{0}\int_{0}^{2\pi}d\varphi_k\int_{0}^{\theta_{0}}d{\theta_k}\sin \theta_k \sqrt{\cos \theta_k} \,
e^{-\gamma_f^2\sin^2\theta_k} \nonumber \\
 && \hspace{60pt} \times \, e^{i\mathbf{k}(\theta_k,\varphi_k)\cdot(\mathbf{r}+\mathbf{r}_s)}\, \mathbf{\hat{x}^{\prime}}(\theta_k, \varphi_k),
\label{E_in}
\eeq
where the direction of the wave vector $\mathbf{k}(\theta_k,\varphi_k)$ is defined by the  angular spherical coordinates $(\theta_k,\varphi_k).$
The region of integration in Fourier space is defined by the semi-aperture angle 
{$\theta_0=\sin^{-1}(\mbox{NA}/n_w)$ if $\mbox{NA}/n_w\le 1$ and $\theta_0=\pi/2$ otherwise.}
All Fourier components share the same modulus $|{\bf k}|=n_w\omega/c$ where $\omega$ is the angular frequency and $c$ is the speed of light in vacuum. 
 We assume that the Gaussian laser beam at the objective entrance port (beam waist $w$) is linearly-polarized along the $x$-direction. 
The unit vector $ \mathbf{\hat{x}^{\prime}}(\theta_k, \varphi_k)$  is  obtained from $\mathbf{\hat{x}}$ by rotation with
 Euler angles $\alpha=\varphi_k,$ $\beta = \theta_k$ and $\gamma=-\varphi_k.$
 We  also define $\gamma_f=f/w$ where $f$ is the objective focal length. 
 Finally, we set the origin at the center of the sphere and the focal point is at position $-\mathbf{r}_s$. 

We take $\mathbf{E}_{\rm in}(\mathbf{r}) $ as the incident field on the the spherical particle to be trapped. 
The Mie-Debye theory of optical tweezers~\cite{MaiaNeto2000,Mazolli2003,Dutra2007} combines Mie scattering
with the Debye-type non-paraxial integral representation (\ref{RW}) for the focused trapping beam. 
Each Fourier component of ${\bf{E}}_{\rm in}$ is scattered by the particle 
 according to the Mie formalism~\cite{Bohren}. The resulting scattered field 
can be written with the help of Wigner finite rotation matrix elements  $d^{\ell}_{m, m'}(\theta)$  \cite{Edmonds}.
We then compute the optical force  ${\bf F}$ by integrating the Maxwell stress tensor over a spherical Gaussian surface $S(r)$ of  radius
 $r\rightarrow \infty:$
  \begin{eqnarray}
 {\bf F}  =  \lim_{r\rightarrow\infty}\left[ -\frac{ r}{2} \int_{S(r)} \! d\Omega \, {\bf r} \left( \epsilon_0\epsilon_w E^2 + \mu_0 H^2 \right)\right].
 \end{eqnarray}
 Here, ${\bf E}$ and ${\bf H}$ are the total electric and magnetic fields, and $\epsilon_0$ and $\mu_0$ denote the vacuum permittivity and permeability, respectively.
 
 After concluding the formal derivation, we displace the origin to the focal point and represent the position of the sphere center  relative to the focus
 $\mathbf{r}_s(\rho,\phi,z)$ in cylindrical coordinates.
 For sake of  convenience, we define  the normalized, dimensionless optical force efficiency \cite{Ashkin1992}
\[
{\bf Q}(\rho,\phi,z)  = \frac{{\bf F}(\rho,\phi,z) }{n_w P/c},
\]
where $P$ is the laser beam power at the sample region.
The axial (${ Q_z}$) and radial (${Q_\rho}$) cylindrical components are of particular interest. The former provides information 
on trap stability along the $z$ axis, which is usually the most difficult direction due to the detrimental effect of radiation pressure. {  To this end, the trapping of {a} passive sphere occurs due to { the interplay between the} {gradient and radiation pressure forces}, where the former  {points} towards the beam focus  {while the latter points}  along 
{ the positive laser beam axis regardless of the particle's position.} Since the  radiation pressure along {any} radial direction is negligible, {the larger gradient force always provides stable {trapping} on the $xy$ plane.}}
However, when considering a particle with gain, { photons
are emitted along one of the directions defined by the Fourier decomposition of the incident beam, as given by (\ref{RW}). If the particle is slightly displaced off-axis, 
the recoil effect provides an outward force component on the $xy$ plane. Thus, the} stability on the $xy$ plane might also be in question when considering gain. Hence it is important to calculate 
the transverse stiffness ($z_{\rm eq}=$ equilibrium position)
 \begin{eqnarray}\label{def_kappa}
 \kappa_\rho= -\frac{n_w\, P}{c}\, \frac{\partial Q_\rho}{\partial \rho}\bigm|^{z=z_{\rm eq}}_{\rho=0},
  \end{eqnarray}
as stable three-dimensional trapping is only achieved when the condition $\kappa_\rho > 0$ is also met in addition to the 
requirement of axial trap stability defined by ${Q_z}.$
We take the $x$ direction corresponding to 
the linear polarization of the laser beam at the objective entrance port when 
calculating the derivative in Eq.~\ref{def_kappa}, 
 since it corresponds to the direction of weaker confinement on the $xy$ plane 
 as the focused spot is elongated along the incident polarization direction~\cite{RichardsWolf}. 

\subsection{ Optical gain model}
{Optical gain media may be achieved for instance by doping the dielectric media with dye molecules (e.g. rhodamine) or solid-state emitters, such as quantum dots or  nitrogen vacancy centers in diamond.  For our purposes,  }
we take the vacuum wavelength $\lambda = 1.064\,\mu{\rm m}$
 and  $n_w=1.332$ for the refractive index of the aqueous medium { hosting the microspheres}. 
 The microsphere is filled by 
 an active medium of  complex refractive index 
 $n_p=\mbox{Re}(n_p)+ i\,\mbox{Im}(n_p)$ with $\mbox{Im}(n_p)<0.$ 
Such condition could be implemented by doping a dielectric material with 
 dye molecules or quantum dots.
 For instance, 
 one can dope a dielectric medium of relative permittivity $\epsilon_h$ with
 four-level dye molecules. The resulting relative permittivity can be modeled by the dispersion relation~\cite{Polimeno2020,Pezzi2019,Campione2011}
\begin{equation}{
 \epsilon_p(\omega) = \epsilon_h(\omega) - \frac{N_{\rm dye}\, \mu^2 }{3\hbar \epsilon_0}\,\frac{ \tilde{N}}{(\omega-\omega_a)+ i /\tau},} \label{gain_media}
\end{equation}
where $\hbar$ denotes the reduced Planck constant.
The dye concentration is represented by the number density
$N_{\rm dye}$ and $\tilde{N}$ is the population inversion coefficient: $\tilde{N}=0$ when the molecules are in the ground state and $\tilde{N}=1$ for a fully inverted population. The remaining parameters appearing in (\ref{gain_media}) characterize the dye molecule:
 $\tau$ is the  relaxation time,  $\mu$  is the amplitude of the transition dipole moment, and $\omega_a$  is the emission frequency.

\begin{figure}
\centering
\includegraphics[width =5. in]{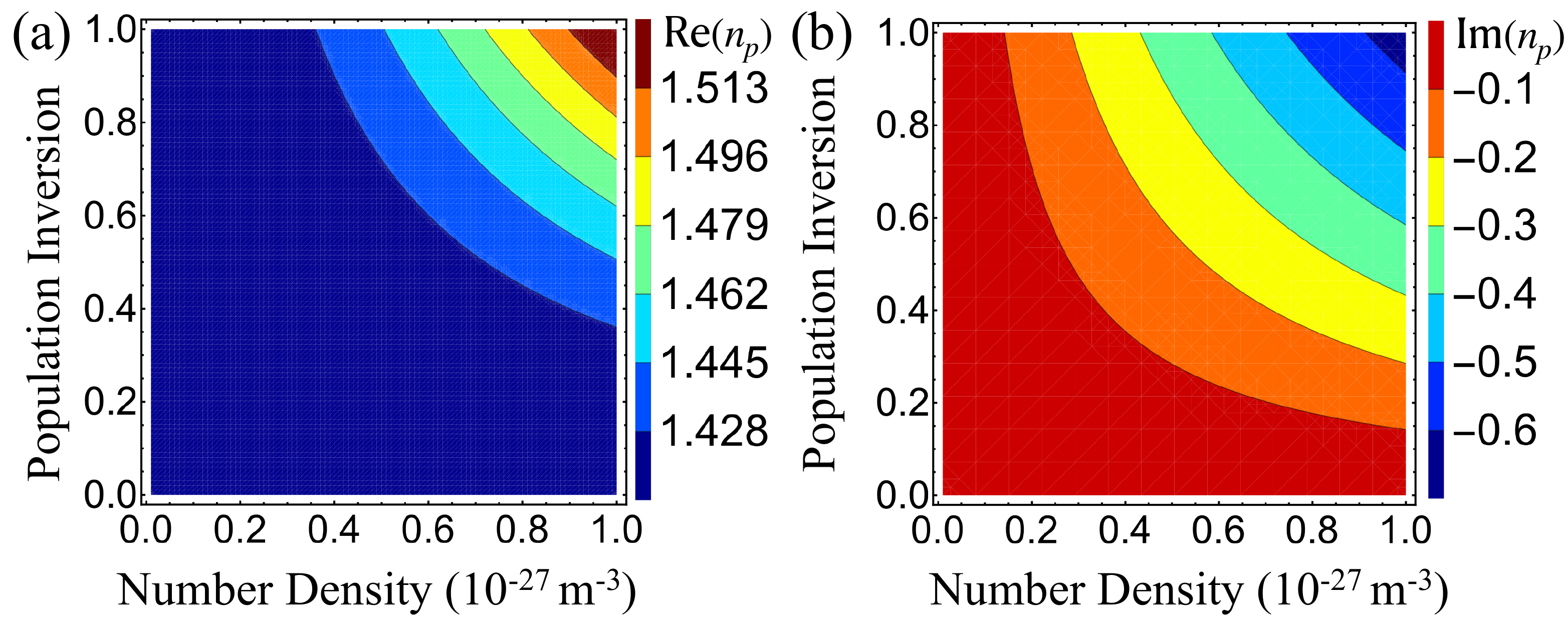}
\caption{ (a) Real and (b) imaginary parts of the  refractive index $ n_p = \sqrt{\epsilon_p} $  of a gain medium
 as functions of the number density of dye molecules $N_{\rm dye}$ and the population inversion $\tilde N$.
 We consider silica as the host medium and take the parameters corresponding to the molecule solvatochromic dye LDS 798~\cite{Doana2017}:  $ \omega_a =  1.77 \times 10^{15}\, {\rm Hz}$,  $\mu = 1.33 \times 10^{-29}\, {\rm C\,m}$ and $ \tau =5 \times10^{-15}\,{\rm s}$.}
 \label{effective_gain}
\end{figure}

 In the numerical examples discussed below, we consider realistic values for the gain level $\mbox{Im}(n_p)<0$
 that can be achieved by choosing the appropriate dye concentration $N_{\rm dye}$ or by
 tuning the power of a pump laser beam driving the population inversion $\tilde{N}.$
 Figure~\ref{effective_gain} shows a density plot of (a) $\mbox{Re}(n_p)$ and (b) $\mbox{Im}(n_p)$ as functions of  $N_{\rm dye}$ and $\tilde{N},$ as calculated from \ref{gain_media}.
For concreteness, we take the example of a silica host medium doped with the molecule solvatochromic dye LDS 798~\cite{Doana2017}, whose parameters are given by Refs.~\cite{Polimeno2020,Pezzi2019}.

\section{ Results and discussion }

\begin{figure}
\centering
\includegraphics[width =4. in]{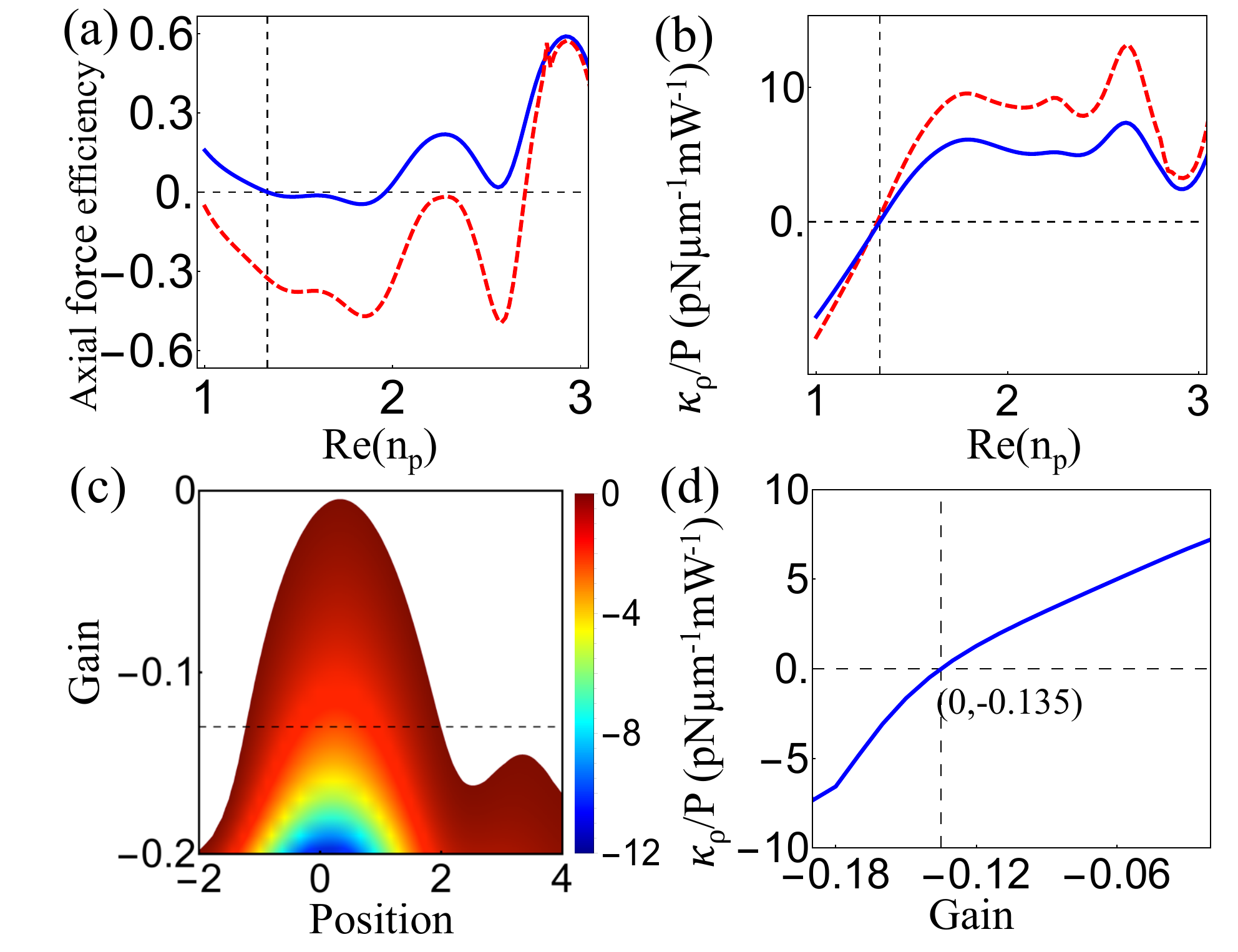}
\caption{Optical force on a homogenous gain-enriched microsphere. (a) Normalized axial force $Q_z$ and (b) transverse trap stiffness  $\kappa_\rho$ (divided by the local power) 
as functions of the real part of the microsphere's refractive index: passive absorptionless (solid) and 
gain functionalized (dash) with Im$(n_p) = -0.05.$ The vertical lines indicate the refractive index  $n_w=1.332$ of the aqueous immersion medium. 
 (c) Density plot of $Q_z$ as function of position (in units of radius) along the beam axis and
 of Im$(n_p)$
  for Re$({ n_p}) = 2.6.$
  The horizontal dashed line indicates the critical value  Im$(n_p)=-0.135.$
  Only negative values of $Q_z,$  which correspond to a backward force, are shown. 
 (d)  Transverse trap stiffness $\kappa_{\rho}$ as function of Im${ (n_p)},$ again with Re${ (n_p) = 2.6}$. The vertical dashed line indicates
 the critical gain Im$(n_p)=-0.135,$ above which stability on the $xy$ plane is lost.} 
 \label{F1}
\end{figure}

In Fig. \ref{F1}, we consider a homogenous microsphere of radius  $R=400\,{\rm nm}$ and typical values for the 
 parameters $\mbox{NA}=1.4$ and $\gamma_f=0.667$ \cite{Dutra2014}.
In panel (a), 
we plot the axial force efficiency $Q_z$  as a function of $\mbox{Re}(n_p)$
 for $\mbox{Im}(n_p)=0$  (solid line), corresponding to 
 a passive, absorptionless particle, 
 and for an active sphere with $\mbox{Im}(n_p) = - 0.05$ (dashed line). We take a fixed position along the beam axis $z/R = 0.2.$
  We represent the value $n_w=1.332$ by a vertical dashed line. 
 
 The solid line in Fig.~\ref{F1}(a) reveals that for passive spheres 
optical tweezing can only occur for a narrow range of refractive indexes, namely $ n_w< \mbox{Re}(n_p) \lesssim 1.92,$
  since axial trapping requires a restoring negative axial force. 
  Indeed, when $ \mbox{Re}(n_p)< n_w $ the polarizability becomes negative and then the particle is expelled from the focal region. On the other hand, for 
  large indexes outside the trapping range, radiation pressure becomes dominant as reflection overcomes refraction, which also prevents trapping. 
  
  However,  trapping is facilitated when the microsphere with the same $\mbox{Re}(n_p)$  and radius is doped with a gain material.
  As indicated by the dashed line in Fig.~\ref{F1}(a), trapping can now be achieved over a much broader range of refractive indexes, as indicated by the 
  range leading to negative optical forces. 
  This is a consequence of the recoil arising from stimulated emission along directions close to the positive $z$ direction, thus leading to a net backward force.
   
 In addition to the stability along the $z$ axis, we also investigate the transverse trap stiffness on the $xy$ plane to verify
 three-dimensional stability of the optical trap.
  We plot $\kappa_\rho$ as a function of Re$({n_p})$ in Fig.~\ref{F1}(b), with the same conventions employed in panel (a).
   The plot shows that the equilibrium position along the axis is unstable when the refractive index is lower than the refractive index of the surrounding medium (dashed vertical line), 
   corresponding to a negative  polarizability.  
   Thus, there is no three-dimensional trapping in this case even when one considers gain. 
    On the other hand, the presence of gain indeed allows for trapping of high refractive index particles. Indeed, as $\kappa_{\rho}>0$ in this case,
    the maximum value of  Re$({n_p})$ allowing for trapping is solely determined by the condition for stability along the axis, which was
     discussed in connection with panel (a). 
\begin{figure}
\centering
\includegraphics[width =5.5 in]{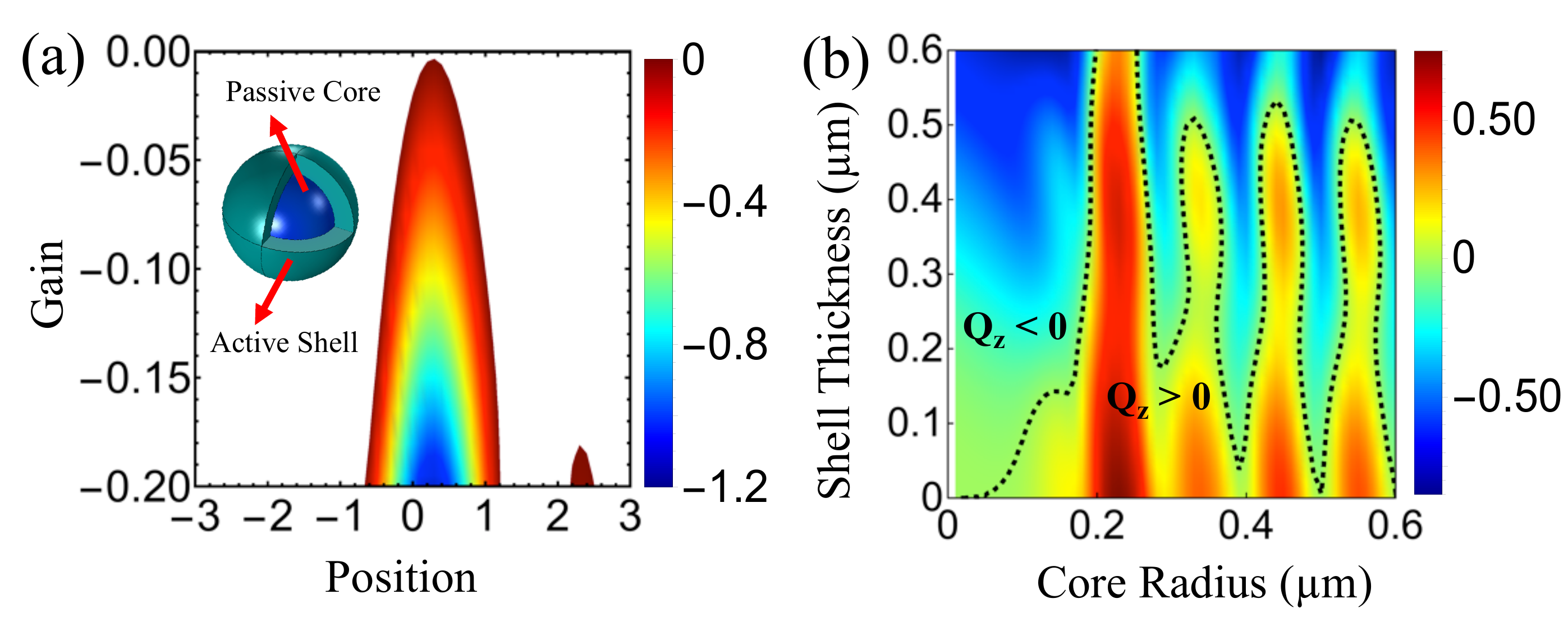}
\caption{ Optical force on a core-shell particle with a dielectric core and a gain-enriched shell. 
(a) Density plot of the normalized axial force efficiency $Q_z$
as a function of axial position (in units of outer radius) and gain. Only negative values are indicated.
The core radius is $R_c=400\,{\rm nm}$ and the shell thickness is $t=200\,{\rm nm}.$
 (b) Density plot of $Q_z$ as a function of  $R_c$ and $t.$ We fix the position at $z/(R_c+t)= 0.1$ and
consider the gain parameter Im$(n_p)=-0.05.$ The dotted line corresponds to $Q_z=0.$ }
 \label{F2}
\end{figure}

Next, we unveil the role of  gain in optical tweezing by 
analyzing the variation of the optical force with the gain parameter  Im$({n_p} )$ for a fixed 
 Re$({n_p} )=2.6.$
 In Fig.~\ref{F1}(c),  we plot the 
axial force efficiency
$Q_z$  as a function of position
along the $z$-axis (in units of radius) and of the gain parameter $\mbox{Im}(n_P).$
The plot reveals that a very small amount of gain is sufficient to allow for tweezing of a high-index particle.
The stable equilibrium position as a function of gain is given by the lefthand boundary of the colored region. 
In the case of passive microspheres interacting with aplanatic focused beams,
 the trapping position is always above focus (positive $z$-axis)~\cite{MaiaNeto2000,Mazolli2003}, as it results from the equilibrium between the positive 
radiation pressure and the negative (backward) force component pointing to the focus and responsible for trapping~\cite{Ashkin1992}. 
Here, however, the equilibrium position is located below the focal plane 
(negative $z$) because of the recoil effect leading to a negative force even below focus. 
Figure~\ref{F1}(c) shows that the particle's equilibrium position can be controlled by changing the gain parameter, which can be achieved with a negligible variation 
of $\mbox{Re}(n_P)$
 by changing, for instance,
the pump power driving the population inversion ${\tilde N}$
as indicated in Fig.~\ref{effective_gain}.
 As expected, the trapping position is displaced further below focus as the gain increases.

However, when the gain parameter increases past the threshold value indicated by the horizontal dashed line in Fig.~\ref{F1}(c), 
the equilibrium position becomes unstable on the $xy$ plane. This is indicated by panel (d), 
showing that the transverse 
trap stiffness $\kappa_{\rho}$ (in units of power) goes negative for  Im$({n_p})<-0.135.$ 
Panels (c) and (d) reveal
a transition from a trapping regime to an optical pulling one 
 as the gain increases, with the $z$-axis becoming unstable as the effect of recoil becomes more dominant. 

After demonstrating that gain facilitates optical trapping and manipulation of homogeneous microspheres, 
 we now consider gain functionalized core-shell particles, which can be nanofabricated by chemically encapsulating dye molecules in the shell~\cite{deluca2012}. 
By adding the geometric aspect ratio to the parameter space, 
the core-shell configuration provides a richer variety of applications and physical effects, as discussed in the following. 

 As a realistic example,
 we consider a coated microsphere composed of a dielectric core of refractive index $ n_c  = 2.6$ and a dye-doped nanoshell of refractive index $ n_s = 1.44 + \mbox{Im}(n_s).$ Figure~\ref{F2}(a) presents a density plot of $Q_z$ as a function of axial position and 
 gain parameter Im$(n_s).$ We consider a core of radius $R_c=400\,{\rm nm}$ and a shell of thickness $t=200\,{\rm nm}.$ 
 Only the regions with negative values of $Q_z,$ allowing for trapping, are shown. As in the case of homogeneous microspheres, a very small amount of gain 
 is sufficient to allow for trapping of  particles with a high-index core. In addition, one can not only displace the trapping position downstream by increasing the gain 
 but also { achieve multiple equilibria  \cite{Viana2007}.
Indeed, Fig.~\ref{F2}(a) shows that a second equilibrium position appears
 at $z= 2.2 (R_c+t),$
 which is, however, only marginally stable.} 

In order to elucidate the role of geometric parameters of the core/shell particle, 
Fig.~\ref{F2} (b) presents a density plot of $Q_z$ as a function of the 
core radius $R_c $ and shell thickness $t$ for the fixed axial position $z= 0.1(R_c+t).$ 
We fix  the gain parameter at 
  $\mbox{Im}(n_s)=- 0.05$ and otherwise take 
 the same parameters of \ref{F2}(a). The dotted line, corresponding to $Q_z=0,$
  represents the boundary separating the region allowing for optical tweezing ($Q_z<0$)
  from the one where radiation pressure dominates and precludes trapping. 
  In all cases  with $Q_z <  0$ in Fig.~\ref{F2}, we verify that $\kappa_{\rho}>0$ so that three-dimensional trapping is possible. 
\begin{figure}
\centering
\includegraphics[width =3.4 in]{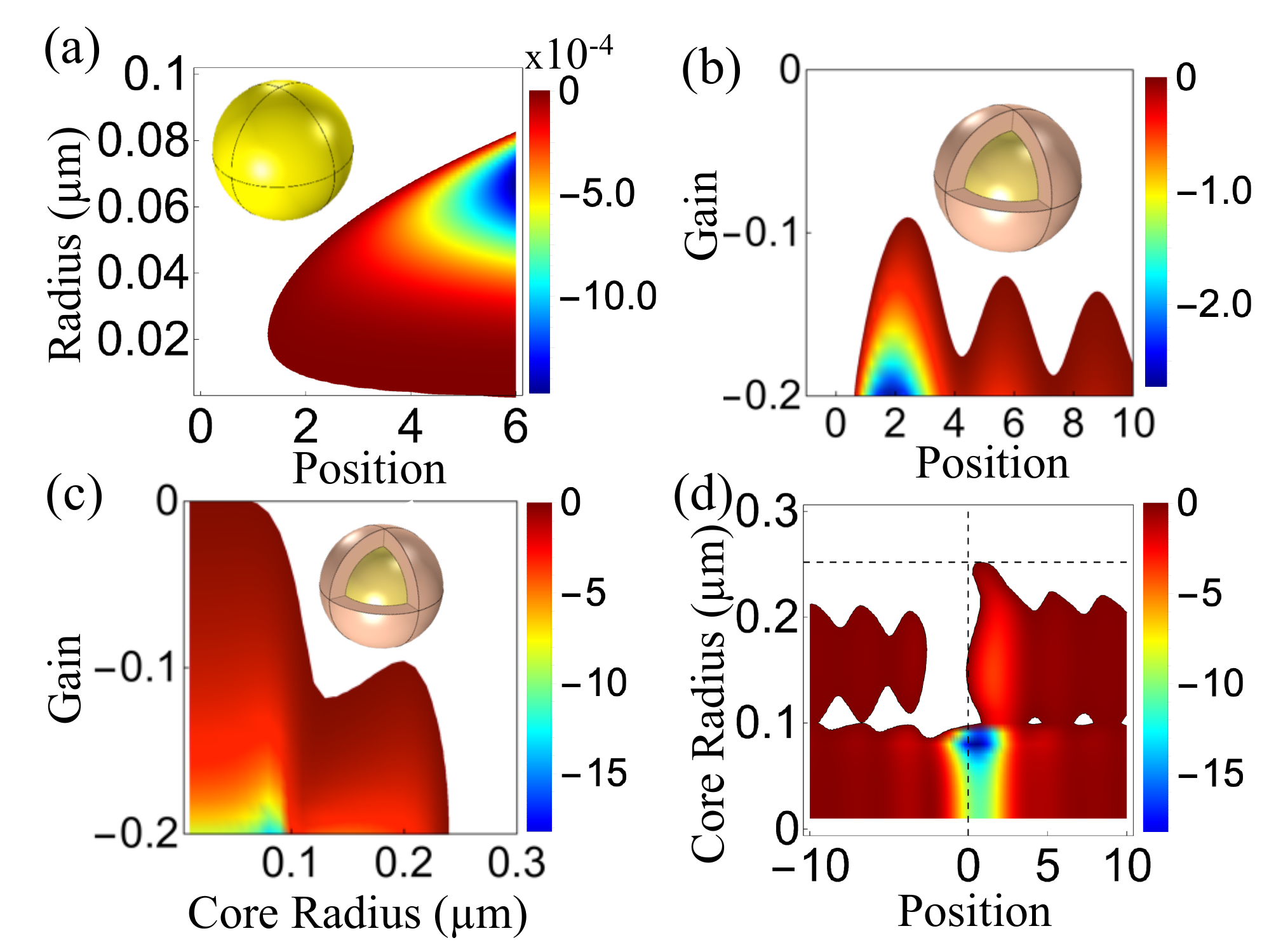}
\caption{Optical force on metallic particles in vacuum. (a) Homogeneous gold particle:
Normalized axial force efficiency $Q_z$ as a function of axial position and radius.
Core-shell particle with gold core and gain-enriched shell: $Q_z$ as a function of (b) axial position and gain parameter
 $\mbox{Im}(n_s),$ with core radius $R_c=200\,{\rm nm}$ 
shell thickness $t= 300\, {\rm nm};$
(c)  core radius and gain parameter  at axial position $z/(R_c+t)=2$; 
(d) axial position and core radius with gain  $\mbox{Im}(n_s)=-0.2.$
In panels (b), (c) and (d), the outer radius is $R_c+t=500\,{\rm nm}.$
Only negative values of $Q_z$ are shown and the axial position is measured in units of the outer radius.
} 
 \label{F3}
\end{figure}

As a third example, we now examine 
the case where the core is metallic, 
as in the nano-fabrication reported by Ref.~\cite{deluca2012}. 
In general, trapping of metallic spheres is difficult due to the dominant role of radiation pressure. 
In addition, strong absorption leads to detrimental radiometric effects when considering an aqueous suspension.
To avoid the presence of thermal effects, from now one we assume that the external medium is vacuum. 
For consistency, we then consider a dry objective with  $\mbox{NA}= 0.95$. 
As an example, we take a homogenous gold nanoparticle with refractive index $n_g=0.25+ 6.96\,i.$ 
We plot the axial efficiency $Q_z$ as function of 
axial position and radius in Fig.~\ref{F3}(a). The panel shows that backward forces and trapping are only achieved for very small particles, in the regime of
validity of the Rayleigh (electric dipole) approximation. 

The size range allowing for trapping of metallic particles can be significantly enlarged by coating the metallic core with a gain medium, as illustrated by panel (b) of 
Fig.~\ref{F3}, showing $Q_z$ as a function of axial position (in units of outer radius) and gain parameter
$\mbox{Im}(n_s)$
 for core radius $R_c= 200\, {\rm nm}$ and shell thickness $t= 300\, {\rm nm}.$ 
Here, the refractive index of the shell is  $n_s = 1.44 + \mbox{Im}(n_s),$
while the refractive index of the gold core is the same as in panel (a). 
The minimum gain  Im$(n_s) \approx -0.09$ required for trapping
is much larger than in the examples considered previously, but still within reach with realistic parameters for the dye doping as indicated by Fig.~\ref{effective_gain}. 
Indeed, in the case of  metallic particles, a larger gain is required to 
compensate for the stronger radiation pressure arising from the enhanced backward scattering and absorption. 

Figure~\ref{F3}(b) shows that 
the equilibrium position can be manipulated by changing the gain, as in the systems considered in Figs.~\ref{F1} and \ref{F2}.
In contrast to the previous configurations, however, equilibrium is achieved above the focal plane because of
 the larger contribution of radiation pressure, and the particle position approaches the focal point from above as the gain increases. 
Multiple stable equilibria appear at intermediate values of the gain parameter. 
As the gain is further increased, 
the separate potential wells merge into a single, deeper well.
More generally, when comparing 
the results of panel (b) with those of panel (a), 
it is worth noticing that the backward negative force increases by roughly four orders of magnitude as the 
plasmonic resonances of the gold core are enhanced by the externally pumped energy associated with gain in the shell region. 

As the metallic core radius $R_c$ increases, so does the detrimental radiation pressure, making it increasingly harder to trap with a single laser beam. 
According to Fig.~\ref{F3}(c), 
showing the variation of $Q_z$ with $R_c$ and  Im$(n_s)$ at the axial position  $z/(R_c+t)=2$ for $R_c+t=500\,{\rm nm},$
 the largest core  allowing for trapping with this arrangement corresponds to  $R_c = 252\,{\rm nm}$
 at the gain parameter  $\mbox{Im}(n_s) = -0.2.$ 
 
 We analyze how the core size modifies the optical force along the $z$-axis for such value of gain 
 in
Fig.~\ref{F3}(d). 
As the core radius decreases from the maximum value $R_c = 252\,{\rm nm},$
the range of positions allowing for a restoring backward force expands and develops multiple equilibria for radii slightly above $0.2\,\mu{\rm m}.$
The plot shows that multiple equilibria also appear below focus (negative $z$) as the radius is further reduced, and then a pulling range develops since the radiation pressure
associated to the core becomes increasingly subdominant in comparison with the recoil effect. 
Indeed, Fig.~\ref{F3}(d)  reveals a 
crossover between optical tweezing and optical pulling~\cite{chen2011} as $R_c$ decreases. Both effects are characterized by a negative force,
but the latter corresponds to a pulling towards the laser source without defining 
 any equilibrium position. 
Such pulling range is first confined to the region below focus for $R_c\approx 0.15\,\mu{\rm m},$ 
and then expands into the entire space as the core radius is reduced below $0.1\,\mu{\rm m}.$

In contrast to the case of homogenous dielectric microspheres discussed in connection with Fig.~\ref{F1}, here the pulling regime preserves stability on the $xy$-plane, 
since the transverse stiffness $\kappa_{\rho}$ remais positive in all cases leading to negative forces in Fig.~\ref{F3}. Thus, as the particle is pulled towards the  laser source, it is also attracted to the beam symmetry axis. 

 \begin{table}[t]
\centering
\begin{tabular}{ |p{3.cm}|p{3 cm}|p{3.cm}|p{3cm}|  }
  \hline
System & Pulling force & Multiple stable points & Minimum gain for {trapping}  \\ 
  \hline
Homogeneous sphere {with gain}  &No&No&$10^{-3}$ \\ 
 \hline
Dielectric core / gain shell&No& Marginal & $10^{-3}$\\ 
  \hline
Metal core / gain shell  & Yes & Yes&$10^{-1}$ \\ 
  \hline

\end{tabular}
\caption{{ Table summarizing the main functionalities that may be achieved with the different gain nanoparticles discussed in this work. }} \label{t1}
\end{table}

{\section{Conclusions}} 
In conclusion, we present  a systematic study of  optical tweezing in the presence of gain functionalized microspheres using a single focused beam. We consider several realistic examples of practical interest including gain enriched homogeneous microspheres and hybrid core/shell plasmonic nanoparticles. We demonstrate that the presence of gain allows for optical trapping in a wide variety of systems that usually cannot be employed in optical tweezers, including plasmonic and large refractive index particles. 
In these cases we show that the trapping conditions strongly depend of the gain value, including the equilibrium position of the trapped  particle, which could be externally controlled  by varying the pump power at realistic conditions. We also demonstrate a crossover between pulling forces and optical forces leading to stable optical tweezing, {as summarized in Table~1.} Altogether, our findings 
 set the grounds for the incorporation of gain in optical tweezers, expanding their employment to new systems such as plasmonic and large, high refractive microparticles, with many potential applications such as self-assembling and gain-assisted optomechanics. 

\section{Acknowledgements}
We thank S. Iqbal, G. Wiederhecker and F. S. S. da Rosa for inspiring discussions. 
 We acknowledge funding from the Brazilian agencies 
 Conselho Nacional de Desenvolvimento Cient\'{\i}fico e Tecnol\'ogico (CNPq),
Coordena\c c\~ao de Aperfei\c coamento de Pessoal de N\'{\i}vel Superior
 (CAPES),  Instituto Nacional de Ci\^encia e Tecnologia de Fluidos Complexos (INCT-FCx),
Funda\c c\~ao de Amparo \`a Pesquisa do Estado do Rio de Janeiro (FAPERJ) (202.874/2017 and 210.242/2018) and
Funda\c c\~ao de Amparo \`a Pesquisa do Estado de S\~ao Paulo  (FAPESP) (2014/50983-3 and 2020/03131-2). 

\end{document}